\documentclass[twocolumn,showpacs,amsmath,amssymb]{revtex4}
\usepackage{latexsym}
\usepackage{amssymb}
\usepackage{amsfonts}
\usepackage{amsmath}
\usepackage[dvips]{graphicx}
\usepackage{bm}

\begin{document}
\title{A post-Keplerian parameter to test gravito-magnetic effects in binary
pulsar systems}
\author{Matteo Luca Ruggiero$^{1,2 }$ and Angelo Tartaglia$^{1,2 }$}
\affiliation{$^{1}$ Dipartimento di Fisica, Politecnico di Torino,
                    Corso Duca degli Abruzzi 23, Torino, Italy\\
             $^{2}$ INFN, Sezione di Torino, Via Pietro Giuria 1, Torino, Italy}

\date{\today}

\begin{abstract}
We study the pulsar timing, focusing on the time delay induced by
the gravitational field of the binary systems. In particular, we
study the gravito-magnetic correction to the Shapiro time delay in
terms of Keplerian and post-Keplerian parameters, and we introduce
a new post-Keplerian parameter which is related to the intrinsic
angular momentum of the stars. Furthermore, we evaluate the
magnitude of these effects for the binary pulsar systems known so
far. The expected magnitude is indeed small, but the effect is
important per se.
\end{abstract}

\maketitle

%----------------------------Section-------------------------------
\section{Introduction}\label{sec:intro}
%----------------------------Section-------------------------------

The first binary radio pulsar (PSR B 1913+16) was discovered some 30 years
ago by Taylor and Hulse\cite{hulse75}. Since those years, a great amount of
work has been done, and pulsars in binary systems have proved to be
celestial laboratories for testing the relativistic theories of gravity (see
\cite{stairs03} and references therein). Indeed, up to the present day,
Einstein's theory of gravity, General Relativity (GR), has passed all
observational tests with excellent results. However, even if the aim of the
experimental relativists is always to achieve a greater precision, we must
not forget that most of the tests of GR come from solar system experiments,
where the gravitational field is in the "weak" regime. On the other hand, it
is expected that deviations from GR can occur for the first time in the
"strong" field regime: hence, the solar system experiments are inadequate to
this end. On the contrary, the strong gravitational field is best tested by
means of pulsars. Pulsars, which are highly magnetized rotating neutron
stars, are important both for testing relativistic theories of gravity and
for studying the interstellar medium, stars, binary systems and their
evolution, plasma physics in extreme conditions. As for the tests of
gravity, the recent discovery of the first double pulsar PSR J0737-3039 \cite%
{burgay03},\cite{lyne04} provided an astonishing quantity of data, which
make this system a rare relativistic laboratory \cite{kramer05}.

As we pointed out elsewhere \cite{tartaglia05}, this system in particular,
and binary pulsars systems in general, could be useful for testing the so
called gravito-magnetic effects. These effects are originated by the
rotation of the sources of the gravitational field, which gives rise to the
presence of off-diagonal $g_{0i}$ terms in the metric tensor. The
gravitational coupling with the angular momentum of the source is indeed
much weaker than the coupling with mass alone, the so called
gravito-electric interaction. In fact, the ratio between the former and the
latter can be estimated to be of the order of
\begin{equation}
\varepsilon =\frac{g_{0\phi }}{g_{00}} \simeq \frac{J R_{S}}{r^{2} Mc}\;,
\label{rapporto}
\end{equation}
where $R_{S}=2GM/c^2$ is the Schwarzschild radius of the source, $M$ being
its mass, and $J$ (the absolute value of) its angular momentum. At the
surface of the Sun, which is the most favorable place in the solar system,
evaluation of eq.~(\ref{rapporto}) gives $\varepsilon \sim 10^{-12}$, thus
evidencing the weakness of the gravito-magnetic versus the gravito-electric
interaction. The smallness of $\varepsilon $ is the reason why, though
having been suggested from the very beginning of the relativistic age~\cite%
{lense}, the experimental verification of the existence of gravito-magnetic
effects has been very difficult until today (see \cite{ruggiero02} and
references therein). The relevance of pulsar systems for the detection of
the gravito-magnetic effects lays in the fact that the ratio~(\ref{rapporto}%
) can be less unfavorable whenever $r$ is approaching the Schwarzschild
radius of the source: this can be the case of a source of electromagnetic
(e.m.) signals orbiting around a compact, collapsed object.

As for the rotation effects in pulsars binary systems, in general, the
coupling of the intrinsic angular momentum of the stars with the orbital
angular momentum, and the coupling of the intrinsic angular momenta of the
two stars themselves were studied, together with the corresponding
precession effects \cite{barker75},\cite{barker76},\cite{oconnell04},\cite%
{weisberg02},\cite{stairs04}. The gravitomagnetic effects on the
propagation of light in a binary system were studied by Kopeikin
and Mashhoon (see \cite{kopeikin02} and references therein).

In a previous paper \cite{tartaglia05} we studied the effects of
the gravitational field on e.m. pulses propagating in a binary
system, and we emphasized the gravito-magnetic contribution. In
doing that, we used a simplified model which considered circular
orbits only. Here we generalize those results, by taking into
account orbits with arbitrary eccentricity, which is a more
realistic approach on the basis of the knowledge of binary systems
discovered until today. In particular, we study the
gravito-magnetic correction to the Shapiro contribution to the
time of flight of the signals, focusing on its effect on the
arrival times perceived by the experimenter on the Earth, and we
introduce a new Post-Keplerian parameter which is related to the
intrinsic angular momentum of the stars. Finally, we evaluate the
magnitude of these effects for the binary pulsar systems known so
far.

%----------------------------Section-------------------------------

\section{The Context: Pulsars Timing}\label{sec:timing}
%----------------------------Section-------------------------------

When studying pulsars, what is measured are the pulse arrival times at the
(radio) telescope over a suitably long period of time. In fact, even though
individual pulses are generally weak and have an irregular profile, a
regular mean profile is obtained by averaging the received pulses over a
long time. Pulses traveling to the Earth are delayed because of the
dispersion caused by the interstellar medium. Besides this delay, other
gravitational factors influence the arrival times of the signals emitted by
a pulsar in a binary system: the strong field in the vicinity of the pulsar,
the relatively weak field between the two compact objects forming the binary
system and, finally, the weak field of the Solar System. Consequently, the
arrival time $T_{N}$ of the $N$th pulse, as measured on the Earth, depends
on a set of parameters $\alpha _{1},\alpha _{2}...,\alpha _{K}$, which
include a description of the orbit of the binary system:
\begin{equation}
T_{N}=F(N,\alpha _{1},\alpha _{2},...,\alpha _{K})  \label{eq:toa1}
\end{equation}%
In particular, the set $\alpha _{1},\alpha _{2}...,\alpha _{K}$ includes the
Keplerian parameters, together with the so called "post-Keplerian" (PK)
parameters which describe the relativistic corrections to the Keplerian
orbit of the system. Since seven parameters are needed to completely
describe the dynamics of the binary system (see \cite{taylor89} and
references therein), the measurement of any two PK parameters, besides the
five Keplerian ones, allows to predict the remaining PK parameters. For
instance, if the two masses are the only free parameters, the measurement of
three or more PK parameter{\normalsize s} over-constrains the system and
introduces theory-dependent lines in a mass-mass diagram that should
intersect, in principle, in a single point \cite{damour91}. This is of
course true as far as the intrinsic angular momenta are not taken into
account.

It is possible to obtain a relation which links the time of arrival of a
pulse on the Earth to its time of emission. More in detail, the following
timing formula holds, which relates the reception (topocentric) time $%
T_{Earth}$ on the Earth with the emission time $T_{pulsar}$ in the comoving
pulsar frame \cite{straumann04}:
\begin{align}
T_{pulsar}=&T_{Earth}-t_0-\frac{D}{f^2}+\Delta_{R_\odot}+\Delta_{E_\odot}-%
\Delta_{S_\odot}  \notag \\
&- \left(\Delta_R+\Delta_E+\Delta_S+\Delta_A \right)  \label{eq:timing1}
\end{align}
where $t_0$ is a reference epoch, $D/f^2$ is the dispersive delay (as a
function of the frequency of the pulses, $f$), $\Delta_R,\Delta_E,\Delta_S$
are, respectively the Roemer delay, the Einstein delay and the Shapiro delay
due to the gravitational field of the binary system (while $%
\Delta_{R_\odot},\Delta_{E_\odot},\Delta_{S_\odot}$ are the corresponding
terms due to the Solar System field) and $\Delta_A$ is the delay due to
aberration.

Here we are concerned with the gravito-magnetic corrections due to the
intrinsic angular momentum of the stars to the Shapiro delay $\Delta_{S}$,
which we analyze in the following section.

%----------------------------Section-------------------------------
\section{The Shapiro Time Delay} \label{sec:shapiro}
%----------------------------Section-------------------------------

%----------------------------Figure-------------------------------
\begin{figure}[tp]
\begin{center}
\includegraphics[width=9cm,height=9cm]{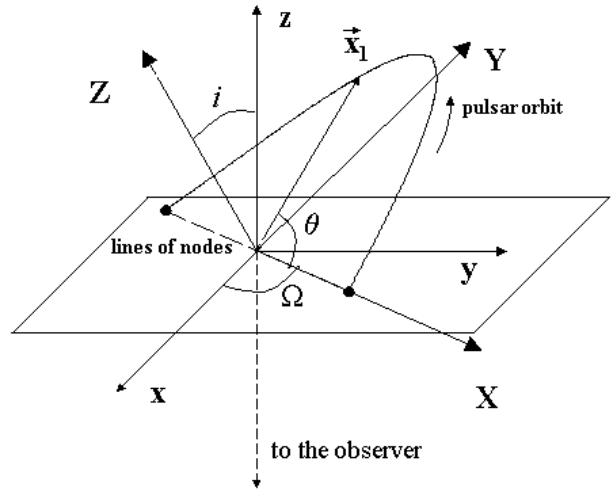}
\end{center}
\caption{{\protect\small Notation used for describing the pulsar orbit.}}
\label{fig:fig1}
\end{figure}
{\normalsize
%----------------------------Figure-------------------------------
}

We want to calculate the relation between the coordinate
emission time $t_e$ and the coordinate arrival time $t_a$ (as measured at
the Solar System center of mass).

To this end, we consider a reference frame at rest in the
center of mass of the binary system. In this reference frame, the vector
pointing to the pulsar emitting e.m. signals is $\vec{\bm{x}}_1$, while the
one pointing to its companion star is $\vec{\bm{x}}_2$; in the following,
the suffix "1" will always refer to the pulsar, and "2" to its companion.
Furthermore, $\vec{\bm{x}}_b$, is the position of the center of mass of the
solar system.

We use the notation of Figure \ref{fig:fig1} for the description
of the pulsar orbit. We choose a first set of Cartesian
coordinates $\{x,y,z\}$, with origin in the center of mass of the
binary system, and such that the line of sight is parallel to the
$z$ axis. Then, we introduce another set of Cartesian coordinates
$\{X,Y,Z\}$, with the same origin: the $X$ axis is directed along
the ascending node, the $Z$ axis is perpendicular to the orbital
plane. The angle between the $x$ and $X$ axes is $\Omega$, the
longitude of the ascending node, while the angle between
the $z$ and $Z$ axes is $i$, the inclination of the orbital plane. Let $\vec{%
\bm{x}}_1=r_1 \hat{\bm{x}}_1$ be the orbit of the pulsar: it is described by
\begin{equation}
\hat{\bm{x}}_1 = \cos(\omega+\varphi) \hat{\bm{X}}+\sin(\omega+\varphi) \hat{%
\bm{Y}},  \label{eq:orbita1}
\end{equation}
in terms of the argument of the periastron, $\omega$, and the true anomaly, $%
\varphi$. Let us pose
\begin{equation}
\theta \doteq \omega+\varphi,  \label{eq:deftheta}
\end{equation}
for the sake of simplicity. Then, we use the notation
\begin{equation}
\vec{\bm{r}} \doteq \vec{\bm{x}}_1-\vec{\bm{x}}_2  \label{eq:defr1}
\end{equation}
to describe the position of the pulsar with respect to its companion, and we
remember that we have, for the Keplerian problem
\begin{equation}
r = \frac{a(1-e^2)}{1+e\cos\varphi},  \label{eq:kep1}
\end{equation}
where $a$ is the semi-major axis of the relative motion and $e$ is the
eccentricity. The astronomical elements $\Omega$, $i$, $\omega$, $a$, $e$
represent the Keplerian parameters. In what follows, we will use also the
definitions $r_b=|\vec{\bm{x}}_b|$, $r=|\vec{\bm{r}}|$ and $\hat{\bm{n}}
\doteq \frac{\vec{\bm{x}}_b}{r_b}$.

That being said, let us focus on the physical situation we are
dealing with. If the gravito-magnetic effects are neglected, the metric
describing the gravitational field of the binary system is given by (see,
for instance, \cite{will})
\begin{equation}
ds^2= \left(1+2\phi \right)dt^2-\left(1-2\phi \right)|d\vec{\bm{x}}|^2,
\label{eq:ds1}
\end{equation}
where
\begin{equation}
\phi (x,y,z) = -\frac{M_1}{|\vec{\bm{x}}-\vec{\bm{x}}_1|}-\frac{M_2}{|\vec{%
\bm{x}}-\vec{\bm{x}}_2|}.  \label{eq:ds2}
\end{equation}
We see that the total gravitational potential is the sum of the two
contributions, due to the pulsar, whose mass is $M_1$, and to its companion,
whose mass is $M_2$\footnote{We use units such that c=G=1; the
signature of the space-time metric is $(1,-1,-1,-1)$.}. Starting from (\ref%
{eq:ds1}), the mass contribution to the time delay can be evaluated
following the standard approach, described, for instance, in \cite%
{straumann04}. However, we want to generalize this approach in order to take
into account the effects of the rotation of the sources of the gravitational
field, i.e. the gravito-magnetic effects.

To this end, we may guess that the metric, in the coordinates $%
\{X,Y,Z\}$, assumes the form
\begin{align}
ds^2= &\left(1+2\phi \right)dt^2-\left(1-2\phi \right)\left(dX^2+dY^2+dZ^2
\right)  \notag \\
&+4 \vec{\bm{A}} \cdot d\vec{\bm{X}}dt,  \label{eq:ds3}
\end{align}
where $\vec{\bm{A}}$ is the (total) gravito-magnetic vector potential of the
system. Eq. (\ref{eq:ds3}) generalizes the weak field metric around a
rotating source (see \cite{mashh1},\cite{mashhoon03}). We suppose that the
dominant contribution to the total gravito-magnetic potential is due to the
intrinsic angular momenta $\vec{\bm{J}}_1,\vec{\bm{J}}_2$ of the stars; we
suppose also that $\vec{\bm{J}}_1,\vec{\bm{J}}_2$ are aligned with the total
orbital angular momentum $\vec{\bm{L}}$ , i.e. perpendicular to the orbital
plane. Furthermore, since we assume that the signals emitted by pulsar 1
propagate along a straight line, and because the gravito-magnetic coupling
depends on the impact parameter, we may conclude that the only relevant
gravito-magnetic contribution comes from the companion star 2. Consequently,
the metric (\ref{eq:ds3}) becomes
\begin{align}
ds^2 = & \left(1+2\phi \right)dt^2-\left(1-2\phi \right)\left(dX^2+dY^2+dZ^2
\right)  \notag \\
& +4J_2\frac{\left(\vec{\bm{x}}-\vec{\bm{x}}_2 \right)\cdot \hat{\bm{X}}}{|%
\vec{\bm{x}}-\vec{\bm{x}}_2|^3} dY dt-4J_2\frac{\left(\vec{\bm{x}}-\vec{%
\bm{x}}_2 \right)\cdot \hat{\bm{Y}}}{|\vec{\bm{x}}-\vec{\bm{x}}_2|^3} dX dt.
\label{eq:ds4}
\end{align}

In (\ref{eq:ds4}) $|\vec{\bm{x}}|$ varies along the straight
line defined by
\begin{equation}
\vec{\bm{x}}(t)=\vec{\bm{x}}_1(t_e)+\frac{t-t_e}{t_a-t_e}\left(\vec{\bm{x}}%
_b(t_a)-\vec{\bm{x}}_1(t_e) \right).  \label{eq:time21}
\end{equation}
If we set
\begin{equation}
\alpha \doteq \frac{t-t_e}{t_a-t_e},  \label{eq:defalfa1}
\end{equation}
then $\alpha=0$ when $t=t_e$, and $\alpha=1$ when $t=t_a$. Consequently, we may write
\begin{equation}
\vec{\bm{x}}(t)=\vec{\bm{x}}_1(t_e)+\alpha\left(\vec{\bm{x}}_b(t_a)-\vec{%
\bm{x}}_1(t_e) \right).  \label{eq:time2}
\end{equation}
After some straightforward manipulations, the line element (\ref{eq:ds4})
can be written in the form
\begin{equation}
ds^2=g_{tt}dt^2+g_{\alpha\alpha}d\alpha^2+2g_{t\alpha}dtd\alpha,
\label{eq:ds41}
\end{equation}
where
\begin{align}
g_{tt} = & 1-\frac{2M_1}{|\vec{\bm{x}}-\vec{\bm{x}}_1|}-\frac{2M_2}{|\vec{%
\bm{x}}-\vec{\bm{x}}_2|},  \notag \\
g_{\alpha\alpha} = & -\left(1+\frac{2M_1}{|\vec{\bm{x}}-\vec{\bm{x}}_1|}+%
\frac{2M_2}{|\vec{\bm{x}}-\vec{\bm{x}}_2|} \right)\left(r_b^2+r_1^2+2r_1 r_b
\sin i \sin \theta \right),  \label{eq:ds42} \\
g_{t\alpha} = & -2J_2\frac{r r_b \sin i \cos \theta}{|\vec{\bm{x}}-\vec{%
\bm{x}}_2|^3}.  \notag
\end{align}

By setting $ds^2=0$, we easily see that the propagation time is
made of three contributions, up to first order in the masses of the stars,
and in the spin angular momentum:
\begin{equation}
\Delta t= \Delta t_0 + \Delta t_M + \Delta t_J,  \label{eq:deltatitot}
\end{equation}
where
\begin{equation}
\Delta t_0 = \int_0^1 \sqrt{r_b^2+r_1^2+2r_br_1\sin i \sin \theta}d\alpha,
\label{eq:tzero1}
\end{equation}
\begin{equation}
\Delta t_M = \int_0^1 \left(\frac{2M_2}{|\vec{\bm{x}}-\vec{\bm{x}}_2|}
\right)\sqrt{r_b^2+r_1^2+2r_1 r_b \sin i \sin \theta}d\alpha,  \label{eq:tM1}
\end{equation}
\begin{equation}
\Delta t_J = \int_0^1 2J_2\frac{r r_b \sin i \cos \theta}{|\vec{\bm{x}}-\vec{%
\bm{x}}_2|^3} d\alpha.  \label{eq:tJ1}
\end{equation}\\
Let us comment on (\ref{eq:tzero1})-(\ref{eq:tJ1}): the first contribution
is a purely geometric one, and is due to the propagation, in flat space, of
e.m. signals from the pulsar toward the Solar System. The second
contribution represents the time depending part of the total Shapiro delay.
As such it is entirely due to the mass of the companion star; in fact it is
easy to verify that the term containing the mass of the pulsar emitting e.m.
signals gives a contribution which remains constant during the orbital
motion. Finally, the third contribution is due to the gravito-magnetic field
of the companion star. To lowest order, we obtain
\begin{equation}
\Delta t_0 \simeq r_b+r_1\sin i \sin \theta  \label{eq:tzero2}
\end{equation}
\begin{equation}
\Delta t_M \simeq 2M_2 \ln \left[\frac{2r_b}{\hat{\bm{n}}\cdot \vec{\bm{r}}+r%
} \right]  \label{eq:tM2}
\end{equation}
\begin{equation}
\Delta t_J \simeq \frac{2J_2 \sin i \cos \theta}{r}\frac{1-\hat{\bm{n}}\cdot
\hat{\bm{r}}}{\sin^2 i \sin^2 \theta -1}  \label{eq:tJ2}
\end{equation}
The contributions (\ref{eq:tzero2}) and (\ref{eq:tM2}) to the time delay are
in agreement with the standard results (see \cite{straumann04}), while
equation (\ref{eq:tJ2}) gives the gravito-magnetic corrections.

From the time delay (\ref{eq:tM2}) we may extract the
contribution which varies during the orbital motion; if we use the equation
of the orbit (\ref{eq:orbita1}) it can be written in the form
\begin{equation}
\Delta t_M^*=-2M_2\ln \left[\frac{1-\sin i \sin(\omega+\varphi)}{%
1+e\cos\varphi} \right].  \label{eq:dsm7}
\end{equation}

If we introduce the PK Shapiro parameters
\begin{eqnarray}
\mathcal{R} & \doteq & M_2,  \label{eq:shapr} \\
\mathcal{S} & \doteq & \sin i,  \label{eq:shaps}
\end{eqnarray}
the mass (or, gravito-electric) contribution to the time delay becomes
\begin{equation}
\Delta t_M^*=-2\mathcal{R}\ln \left[\frac{1-\mathcal{S} \sin(\omega+\varphi)%
}{1+e\cos\varphi} \right].  \label{eq:dsm8}
\end{equation}

On the other hand, the gravito-magnetic contribution (\ref%
{eq:tJ2}) changes continuously during the orbital motion, so that,
similarly, we may write
\begin{equation}
\Delta t_J^* = -\frac{J_2 \sin i }{a (1-e^2)}\left[\frac{\left(\cos(\omega+%
\varphi)\right)\left(1+e\cos \varphi\right)}{1-\sin i \sin (\omega+\varphi)} %
\right].  \label{eq:dsj3}
\end{equation}
If we introduce the PK parameter $\mathcal{S}$, and define a new PK
parameter
\begin{equation}
\mathcal{J} \doteq J_2  \label{eq:j2p}
\end{equation}
the gravito-magnetic contribution to the Shapiro time delay can be written
in the form
\begin{equation}
\Delta t_J^* = -\frac{\mathcal{J} \mathcal{S} }{a (1-e^2)}\left[\frac{%
\left(\cos(\omega+\varphi)\right)\left(1+e\cos \varphi\right)}{1-\mathcal{S}
\sin (\omega+\varphi)} \right].  \label{eq:dsj4}
\end{equation}
We notice that the new PK parameter $\mathcal{J}$ coincides with the
intrinsic angular momentum of the source of the gravito-magnetic field. In
particular, if we knew the rotation frequency of the source of the
gravito-magnetic field (which is possible, if the latter is a visible
pulsar, for instance), the new PK parameter could give information on its
moment of inertia.

The gravito-electric and gravito-magnetic contributions to the
time delay (\ref{eq:dsm8}) and (\ref{eq:dsj4}) can be written in the form
\begin{eqnarray}
\Delta t_{M}^{\ast } &=&A_{M}F_{M}(\varphi ),  \label{eq:afm1} \\
\Delta t_{J}^{\ast } &=&A_{J}F_{J}(\varphi ),  \label{eq:afj1}
\end{eqnarray}%
where we have introduced the following constant "amplitudes"
\begin{eqnarray}
A_{M} &\doteq &-2\mathcal{R},  \label{eq:afm2} \\
A_{J} &\doteq &-\frac{\mathcal{J}\mathcal{S}}{a(1-e^{2})},  \label{eq:afj2}
\end{eqnarray}%
and the varying "phase" terms
\begin{eqnarray}
F_{M}(\varphi ) &\doteq &\ln \left[ \frac{1-\mathcal{S}\sin (\omega +\varphi
)}{1+e\cos \varphi }\right] ,  \label{eq:afm3} \\
F_{J}(\varphi ) &\doteq &\frac{\left( \cos (\omega +\varphi )\right) \left(
1+e\cos \varphi \right) }{1-\mathcal{S}\sin (\omega +\varphi )}.
\label{eq:afj3}
\end{eqnarray}%
In particular we see that when $\mathcal{S}=\sin i=1$,
$F_{J}(\varphi )$ tends to diverge as far as $\omega +\varphi
\rightarrow \pi /2$ (i.e. close to the conjunction position, when
the impact parameter goes to zero). $F_{M}$ too diverges in the
same configuration and we see that the divergences have different
"strength", the former being an inverse power, the latter
logarithmic. The reason for that difference is easily referable to
the gravito-magnetic potential affecting  $F_{J}$, which has a
dipolar structure, and to the monopolar gravito-electric one
affecting $F_{M}$.

Even though the divergences have no physical meaning because
the actual compact objects have finite dimensions and the e.m. signals
cannot pass through the center of the companion star, we see that the
gravito-magnetic contribution is bigger for those systems that are seen
nearly edge-on from the Earth, which are the ideal candidates
for revealing the gravito-magnetic effects. This is the case, for instance,
of the binary system PSR J0737-3039, where, however, unfortunately the
presence of a large magnetosheath zone makes the effective impact parameter
much bigger than the actual linear dimension of a neutron star \cite%
{tartaglia05}. That being said, in the following section we give numerical
estimates for the constant amplitudes $A_{M}$ and $A_{J}$ for the known
binary pulsar systems.

%----------------------------Section-------------------------------
\section{Numerical Estimates}\label{sec:numerical}
%----------------------------Section-------------------------------

The PK parameters related to the Shapiro time delay $\mathcal{R}%
,\mathcal{S}$ have been successfully measured with great accuracy in some
binary pulsar systems, such as PSR B 1913+16 (see \cite{weisberg02},\cite%
{taylor89}), and the recently discovered system PSR J0737-3039 (see \cite%
{burgay03},\cite{lyne04},\cite{kramer05}). Indeed, the analysis of these
systems is very accurate, because of their favorable geometrical properties
and, through the measurements of several PK parameters, they provided very
accurate tests of GR as confronted to alternative theories of gravity. In
\cite{tartaglia05}, we studied the gravito-magnetic corrections to pulsar
timing in a simplified situation, taking into account circular orbits only.
Here, since we have generalized those results to arbitrary elliptic orbits,
we may apply the formalism developed so far to all the binary pulsar systems
known up to this moment, in order to estimate the magnitude of the
gravito-magnetic corrections to the time delay. In Table \ref{tab:table1} $%
A_{M}$ and $A_{J}$ are evaluated for the binary systems known until today%
\footnote{Notice that, in physical units we have $A_{M}=\frac{%
GM_{2}}{c^{3}}$, $A_{J}=\frac{GJ_{2}\sin i}{c^{4}a(1-e^{2})}$.}. A few
comments on how the table has been obtained: for those systems where the
available data are not complete, we have chosen for the missing data the
most favorable value (see the caption of the table). Furthermore, we have
estimated the intrinsic angular momentum of the sources of the gravitational
field supposing that the progenitor star was only a little bigger than the
Sun, and that most of the angular momentum was preserved during the
collapse, so that $J_{2}\simeq J_{\odot }$.

From Table \ref{tab:table1} it is clear that the
gravito-magnetic contribution is much smaller than the mass contribution,
as expected. However it is possible, at least in principle, to
distinguish the former from the latter, on the basis of their different
dependance from the geometric parameters. In fact, from (\ref{eq:tzero2}),(%
\ref{eq:dsm8}) and (\ref{eq:dsj4}) it is clear that the geometric and the
gravito-electric contribution are symmetric with respect to the conjunction
and opposition points, while the gravito-magnetic contribution is
anti-symmetric. As we pointed out in the previous paper \cite{tartaglia05},
if it were possible to identify conjunction and opposition points in the
sequence of arriving pulses, this fact could be exploited for extracting the
gravito-magnetic effect.

Nowadays, the uncertainties in pulsar data timing are of the
order of $10^{-6}\sim 10^{-7} s$ (see for instance \cite{kramer05}). However, as we
pointed out above, gravito-magnetic effects can become larger if the
geometry of the system is favorable: in particular, when $\sin i=1$, $F_{J}$
tends to diverge close to the conjunction. On the other hand, we can give an
estimate of the value of the geometrical parameters needed to make $A_{J}$
of the order of magnitude of the present day uncertainties. So, if we assume
$J_{2}\simeq J_{\odot }$, in order to have $A_{J}\geq 10^{-7}\ s$, we must
have
\begin{equation}
a(1-e^{2})\leq \frac{a(1-e^{2})}{\sin i}\leq 5\times 10^{4}\ m
\label{eq:estim1}
\end{equation}%
Hence, small orbits with great eccentricity allow, at
least in principle, the measurement of the gravito-magnetic effects. It
might be useful, also, to estimate the rate of decay of the orbit because of
the emission of gravitational waves. If we assume, for the sake of
simplicity, $e=0,M_{1}=M_{2}=1.44M_{\odot }$ and that $a$ fulfills (\ref{eq:estim1}),  we get (see \cite{straumann04}%
)
\begin{equation}
\left\vert \frac{a}{\dot{a}}\right\vert =\frac{5}{64}\frac{a^{4}}{%
M_{1}M_{2}(M_{1}+M_{2})}\leq 2\times 10^{7}\ s  \label{eq:estim2}
\end{equation}%
Consequently, we may argue that these effects may become
larger in the final phase of the evolution of the binary systems, i.e.
during their coalescence, even though, in that phase, the
weak field approach that we used in this paper would probably be rather poor%
demanding for different analysis  techniques.

The smallness of the gravito-magnetic delay would also require long data
taking times so posing the problem of the stability of the pulsar frequency.
However a peculiarity which is not blurred by any drift or noise is the
physical antisymmetry of the gravito-magnetic effect, which should emerge in
the long period over all other phenomena.
%----------------------------Table-------------------------------
\begin{table}[top]
\begin{center}
\medskip
\begin{tabular}{lcc}
System & $A_M$ $[\mu s]$ & $A_J$ $[p s]$ \\ \hline
{\small PSR B1913+16$^{a}$} & 6.9 & 4.2 \\
PSR J0737-3039$^{b}$ & 6.2 & 11.8 \\
PSR B1534+12$^{c}$ & 6.7 & 2.3 \\
PSR J1756-2251$^{d}$ & 5.9 & 2.8 \\
PSR J1829+2456$^{e}$ & 6.1 & 1.1 \\
PSR J1518+4904$^{f}$ & 7.2 & 0.5 \\
PSR J1811-1736$^{g}$ & 3.5 & 0.4 \\
PSR B2127+11C$^{h}$ & 6.8 & 6.1 \\ \hline
&  &  \\
$^a$: \cite{weisberg02},\cite{taylor89} &  &  \\
$^b$: \cite{kramer05} &  &  \\
$^c$: \cite{stairs02},\cite{stairs04},\cite{thorsett04} &  &  \\
$^d$: \cite{faulkner04} &  &  \\
$^e$: \cite{champion04} &  &  \\
$^f$: \cite{nice96} &  &  \\
$^g$: \cite{lyne99} &  &  \\
$^h$: \cite{deich96} &  &  \\
&  &
\end{tabular}
\end{center}
\caption{\small Evaluation of the gravito-electric and
gravito-magnetic contributions to the time delay. For the systems PSR
J1756-2251,PSR J1829+2456, PSR J1518+4904, PSR J1811-1736, PSR B2127+11C,
since the present data do not provide the inclination of the orbit, we chose
the most favorable value for the calculations of $A_J$, i.e. $\sin i=1$.
Similarly, for the calculations of $A_M$, we chose the best estimate for the
mass of the companion star for the systems PSR J1829+2456, PSR J1518+4904,
PSR J1811-1736, since the available data do not constrain it completely.}
\label{tab:table1}
\end{table}
%----------------------------Table-------------------------------
\\

%----------------------------Section-------------------------------
\section{ Conclusions}\label{sec:conc}
%----------------------------Section-------------------------------

In this paper, we have studied the effects of the gravitational
interaction on the time delay of electromagnetic signals coming from a
binary system composed of a radio-pulsar and another compact object. In
particular, we have focused our attention on the gravito-magnetic
corrections to the time delay due to the gravitational field of the binary
system (Shapiro time delay).\newline
\indent In doing so, we have generalized the results obtained in a previous
paper, where we considered a simplified situation, taking into account
circular orbits only. Here arbitrary elliptic orbits are allowed.
Furthermore, by following a standard approach, we have expressed the time
delay and its gravito-magnetic component in terms of Keplerian and
post-Keplerian parameters.\newline
\indent In particular, a new post-Keplerian parameter has been introduced,
which coincides with the intrinsic angular momentum of the source of the
gravitational field, and could give, in some cases, information on its
moment of inertia.\newline
\indent We have given numerical estimates of the amount of the
gravito-magnetic corrections for all binary pulsar systems known until
today, and we have seen that, even though they are usually much smaller than
the corresponding gravito-electric ones, they can become larger for those
systems that are seen nearly edge-on from the Earth, which are the ideal
candidates for revealing the gravito-magnetic effects in this context.%
\newline
\indent Among the known binary pulsar systems, the one having the
most favorable geometrical properties for the detection of the
gravito-magnetic effect is PSR J0737-3039, which, unfortunately,
has a large magnetosheath that keeps the magnitude of the
gravito-magnetic correction below the detectability threshold.
However, we cannot exclude, at the present discovery rate of new
binary pulsars, that other binary systems with favorable
geometrical configurations can be found. This fact, together with
the expected improvement of the sensitivity and precision of the
timing of pulses, makes us cherish the hope that, in the future,
it will be possible to measure the gravito-magnetic corrections to
the time delay, and in particular the newly introduced
post-Keplerian parameter.

%----------------------------Bibliography-------------------------------

%----------------------------Bibliography-------------------------------
\end{document}